\documentclass{article}

\usepackage[colorlinks,urlcolor=blue,linkcolor=blue,citecolor=blue]{hyperref}
\usepackage{color}
\usepackage{authblk,graphicx}

\begin{document}

\title{Using Science Education Gateways to improve undergraduate STEM education: The QUBES Platform as a case study}

\author[1]{Sam Donovan}
\affil[1]{BioQUEST Curriculum Consortium, 5917 Alder St, Pittsburgh, PA, 15232, USA}

\author[2]{M. Drew LaMar}
\affil[2]{William \& Mary, Williamsburg, VA, 23187, USA}

\maketitle

\begin{abstract}\looseness-1The QUBES platform was conceived as a ``science education gateway'' and designed to accelerate innovation in undergraduate STEM education. The technical infrastructure was purpose built to provide more equitable access to professional resources, support learning that reflects authentic science, and promote open education practices. Four platform services (OER Library Access; Professional Learning; Partner Support; and Customizable Workspaces) support overlapping faculty user communities, provide multiple points of entry, and enable manifold use case scenarios. The integrated nature of the platform makes it possible to collect, curate, and disseminate a diverse array of reform resources in a scalable and sustainable manner. We believe that the QUBES platform has the capacity to broaden participation in scholarship around teaching and learning and, furthermore, that it can help to lower faculty barriers to the adoption of reform practices. The role of cyberinfrastructure in undergraduate STEM education is generally underappreciated and warrants further exploration.
\end{abstract}

\section{INTRODUCTION}
The impacts of science gateways on research outputs across diverse fields are well documented \cite{Barker}. Gateway infrastructures can be used to support emerging practices that emphasize the interdisciplinary, collaborative, open, and computationally driven nature of science. However, the adoption of a science gateway as a research platform can require significant adjustments to existing workflows and challenge one’s assumptions about how to manage a research program. Gateways that do not engage their user communities to build trust and support adoption are less likely to establish a robust research community and therefore limit their potential scientific impacts \cite{Kee}. The acceptance and use of scientific gateways, like the diffusion of any innovation, depends in large part on the perceived usefulness of the resource. Developing gateways collaboratively, with input from user communities may accelerate the adoption of new disciplinary practices afforded by the gateway platform. Centralized resources and online collaboration are becoming more widely adopted in many contexts where we believe the ``gateway model'' provides a context for addressing hard problems and broadening access to key resources.

In this manuscript we describe an effort to build a Science Education Gateway to accelerate undergraduate STEM education reform. We define reform very broadly to embrace the diversity and dynamic nature of the landscape across which reform happens. Effective teaching must be supported in a wide range of institutional settings, student populations, and delivery modalities. There is broad recognition that teaching and learning strategies should evolve to emphasize the adoption of evidence-based teaching methods, student engagement with authentic scientific practices, and broaden participation among traditionally underrepresented communities. These challenges are further complicated by the need to continuously integrate new topics and skills to connect classrooms to contemporary scientific practices and help prepare students to participate in the technical workforce. Even when innovations are developed it is a non-trivial undertaking to support the broad implementation of those strategies so that the potential benefits reach as many learners as possible. Considered at this scale, the acceleration of STEM education reform is a wicked problem that will require the development of diverse overlapping strategies that can be applied flexibly across the landscape. Furthermore, given the certainty that scientific practices and computational resources will continue to evolve, education reform should adopt a continuous quality improvement process in order to treat reform as ongoing and context specific. 

In 2014 NSF funded the project ``Supporting Faculty in Quantitative Undergraduate Biology Education and Synthesis (QUBES)'' which was designed to ``address the Nation's growing need to better prepare undergraduate biologists with the quantitative and computational skills needed to be successful in the workplace or in graduate school.'' \cite{Donovan} Given the long history of quantitative biology education reform efforts, the project was organized in part to highlight the visibility of ongoing but isolated reform communities and coordinate faculty access to a diverse collection of existing teaching and learning resources. We adopted the HubZero platform and worked with the Science Gateways Community Institute (SGCI) to design and deploy a gateway to support quantitative biology education innovation and classroom implementation. Over time our mission has evolved to serve the STEM education reform community more broadly. At the conclusion of the initial NSF funding the management of the QUBES platform was moved into the BioQUEST Curriculum Consortium, a well established 501(c)(3) nonprofit, where it is sustained as an open resource for the reform community.

In this paper we describe the conceptualization and implementation of the QUBES platform as a Science Education Gateway (SEG). After an overview of the technical infrastructure (tools) we describe the ways that faculty use of the gateway is facilitated using social infrastructure (practices). We end with a call to action for the undergraduate STEM education reform community to explore the potential use of science education gateways as a means to accelerate the reform of teaching and learning.
\vspace*{-5pt}

\section{THE QUBES PLATFORM AS A SCIENCE EDUCATION GATEWAY}

\begin{figure}
\centerline{\includegraphics[width=18.5pc]{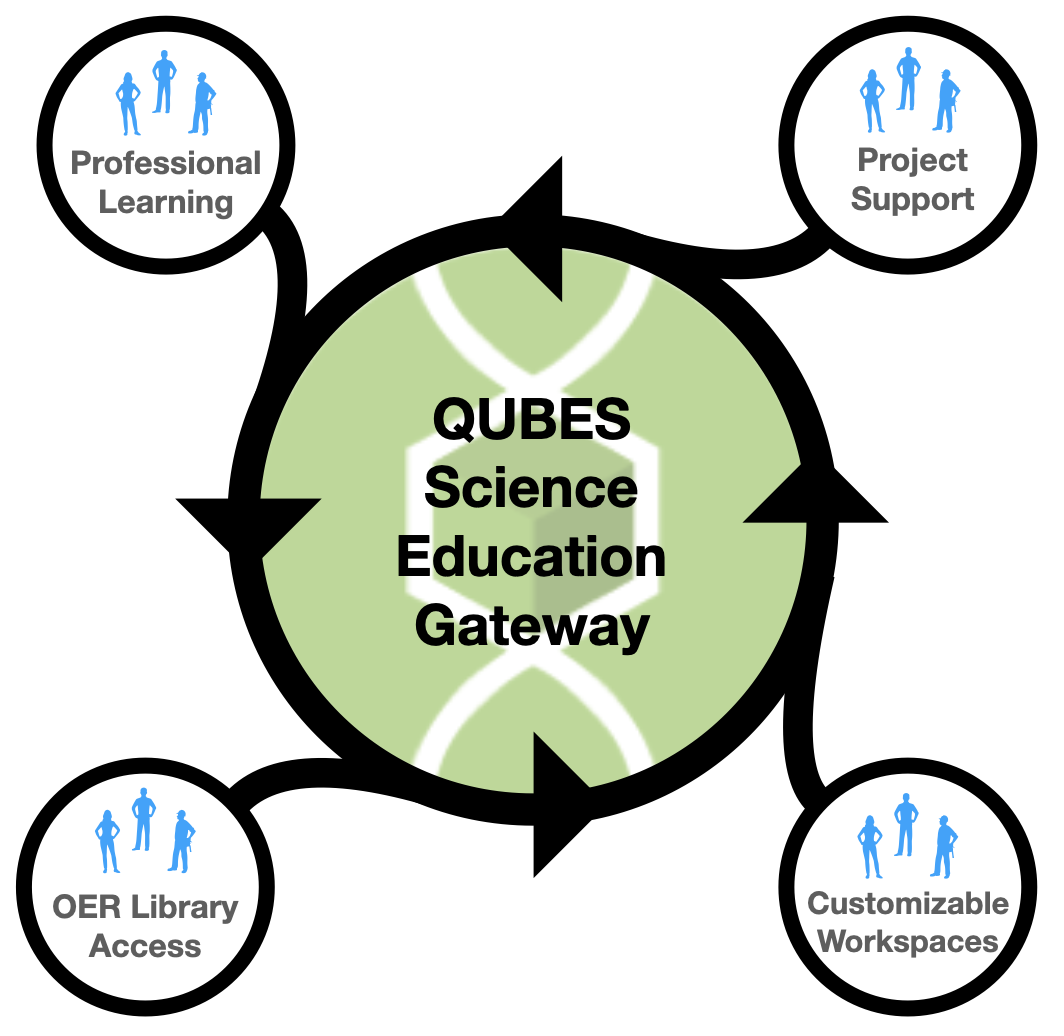}}
\caption{{\bf QUBES Science Education Gateway and its Four Platform Services.} The central element, labeled QUBES Science Education Gateway, represents the underlying technical infrastructure that supports an integrated, customizable online platform. The four platform services (Professional Learning; OER Library Access; Project Support; and Customizable Workspaces) support overlapping faculty user communities, provide multiple points of entry, and enable manifold use case scenarios. Broadly, the gateway can be described as both a platform for hosting diverse professional activities and as a highly curated repository of teaching and learning resources. This dual functionality supports a virtuous cycle where faculty engagement with STEM reform helps them generate new products which are then captured and curated within the gateway, making those resources accessible to the broader community.}
\label{fig:seg}
\vspace*{-5pt}
\end{figure}

Gateways refer to community-developed online environments that integrate access to shared resources including software, data, collaboration tools, and high-performance computing. As a Science Education Gateway QUBES is designed to lower barriers to faculty participation in STEM education reform by making it easier to engage in scholarship around teaching and learning. From finding new teaching materials, to collaborating on projects, to accessing interdisciplinary expertise the QUBES platform provides both technical and social support to engage faculty. Our target audiences include faculty whose scholarship centers on teaching and learning, with the platform designed to facilitate, document, and disseminate faculty work as they participate in diverse professional activities. Following a high level overview of the technical infrastructure, we describe a set of platform services that provide opportunities for faculty to engage with reform resources and pursue professional opportunities (Figure \ref{fig:seg}).
\vspace*{-5pt}

\subsection{Technical Infrastructure}
The QUBES platform is a shared online space that can be used to publish and disseminate Open Education Resources, host distributed meeting and workshop activities, participate in professional learning, and support education reform projects. QUBES is an instantiation of the open source content management system HubZero, which initially was a branch of the Joomla! content management system (CMS) but is now an independently developed platform for scientific gateways. 

The underlying technical infrastructure provides community building tools, including communication, productivity, and collaboration functionality, through membership controlled group spaces. These group spaces can be public or private, with differing levels of openness to new members (e.g., completely open, curated admission, invitation only). There can be multiple administrators assigned to a group who have additional abilities, such as changing group settings, controlling membership, and creating or modifying group web pages. Each group space has optional community tools, such as discussion forums with email digest capabilities, announcements, calendars, file sharing, blogs, Pinterest-style file and image sharing (Collections), collaborative project spaces, a dedicated OER library, and traditional usage and website metrics. Group spaces can be fully customized to have their own web templates allowing for a customized look-and-feel.

The HubZero CMS also supports the publication and hosting of software tools. QUBES has a dedicated execution host for providing cloud-based access to any software that can be run on a linux machine using the OpenVZ middleware. Tools available on QUBES include data science environments (e.g., RStudio Server, Jupyter Notebooks, Shiny apps), modeling tools (e.g., NetLogo, Avida-Ed) and analysis tools (e.g., ImageJ, Mesquite). HubZero is currently working to replace OpenVZ with Docker containers, which would provide additional functionality and better scalability. The QUBES platform currently has the capacity to run 200 concurrent tool sessions which presents significant scaling challenges when making software resources available to a national audience of STEM educators and students.

The QUBES platform has an open, self-publishing platform (QUBES OER Library) that uses a git-like version control system for tracking versions, adaptations, attribution, and use metrics. For searchability, publications are categorized via multiple standard OER library ontologies, such as activity length and audience level. Additional optional ontologies are available to align resources with evidence-based pedagogical frameworks, such as inclusive pedagogy, universal design for learning, and open science and education practices \cite{Dewsbury}--\cite{Cangialosi}. Publications receive a digital object identifier (DOI) and several use metrics are automatically tracked. Importantly, publications can be associated with not just authors, but projects, organizations, or other QUBES hosted groups. This makes it easy to extract and display subsets of the library within a project group context, contributing to partner customization and autonomy.

While QUBES is utilizing the open-source HubZero CMS, we have pushed for the development of additional functionality and UX improvements, both independently and through the support of the SGCI Developer Program. For example, in collaboration with HubZero developers we implemented an email digests option for group discussion forums. Similarly, in collaboration with the SGCI Developer Program we implemented a forking/adaptation system for open education resources that made it possible to support the full OER lifecycle. Independently developed items include supporting instructor-only access to particular files within a published resource, an OER commenting system, and group webpage usage metrics. To support autonomy of hosted projects on the QUBES platform, we increased group customization on the platform by allowing overrides of components, plugins and modules, including fully autonomous OER libraries (e.g., a dedicated search-and-browse for their community resources, and a customized resource record view). Finally, we have developed the ability for OER to have multiple aligned ontologies that can be easily used in a search-and-browse interface with Solr search capabilities.
\vspace*{-5pt}

\section{PLATFORM SERVICES}
In order to raise faculty awareness and encourage adoption of the QUBES platform we have developed a set of four platform services (Figure \ref{fig:seg}) tied to common uses of the underlying technological infrastructure (tools). These platform services promote a set of use scenarios, or practices, which help to contextualize the ways that the gateway technology can be leveraged to meet faculty needs. Platform services use a social infrastructure which provides training, templates, activity structures, and support documents to lower barriers to participation and promote successful engagement with the gateway. Developing and sharing strategies for effectively using a gateway can address both technical challenges associated with manipulating the gateway tools, and introduce new professional practices that incorporate things like open licensing, distributed collaborations, and online community building.\vspace*{-5pt}

\subsection{Open Education Resources Library Access}
\looseness-1Too often innovative educational approaches developed by STEM faculty are not widely shared, licensed to support reuse, or documented as professional scholarship. Our limited ability to capture, curate, and disseminate the collective teaching and learning knowledge of undergraduate STEM faculty severely limits the development and implementation of reform practices. Collections of education resources are often distributed across multiple web sites making them difficult to find and manage, in addition to being more difficult to sustain over time. 

The QUBES OER Library currently hosts over 2,100 resources that range from conference posters to classroom activities. This centralized, well described collection of teaching resources increases findability and can help faculty explore new teaching and learning strategies. By hosting diverse resources across many projects the QUBES OER Library promotes search both within and across projects. Rich contextualization of the product reflecting authorship, project association, group affiliation and other meta-data makes it possible for users to naturally follow threads connecting seemingly disparate materials to support the discovery of related content. This also makes it easy to extract and display subsets of the library within a project group context, contributing to partner customization and autonomy.

Access to the QUBES OER Library makes it easy for faculty to share their own work. Because the library uses open licensing and the infrastructure contains a ``git-like'' version management system it is possible for faculty to engage in the full OER lifecycle. Adaptations of existing resources automatically contain attribution information that link the resources, making it easy for users to navigate between them. Use metrics address a range of data (e.g., views, downloads, comments, versions, and adaptations) that help authors assess the impacts of their products within the community.

The introduction of social infrastructure (practices) around open licensing has been important to address community concerns about OER. Hosted projects on QUBES are often very interested in publishing their work as part of their dissemination and sustainability. We also encourage projects to share non-traditional products such as annual reports, conference presentations, protocols and manuals to document outcomes and make them accessible to the broader community. To support faculty use of the QUBES OER Library we structure professional activities (e.g., project collaborations, professional learning, workshops) around existing published materials, or as a mechanism to collect newly generated resources. This immediately helps to establish an authentic context for participating in professional learning opportunities – run by BioQUEST or hosted partner organizations – and generate products that are appropriate for publication. We have adopted the language of a preprint server to help faculty understand the benefits of sharing works in progress and recognise the utility of self-publishing in the educational space. 

The BioSkills Guide is an example of a resource that was shared early during its development and then versioned as it was refined. To date it is in its fifth version and in total it has been accessed more than 7,500 times and downloaded more than 2,500 times \cite{Clemmons}. When the manuscript describing the development of the guide was published in a peer reviewed journal, the guide on QUBES was referenced, with the journal citation added to the description of the guide on QUBES, thereby pushing discoverability and traffic in both directions.
\vspace*{-5pt}

\subsection{Project Support}
\looseness-1Externally funded education reform projects play an important role in fostering innovation and exploring effective teaching practices. However, funded projects often face logistical challenges like coordinating project activities, documenting project impacts and sustaining their work beyond the funding period. It is essential to the ongoing reform of STEM education that funded projects are as successful as possible and that their findings are carefully documented and shared to increase their impact over time. Effectively engaging user communities early in a project’s development cycle can play an important role in guiding the work to be broadly useful and seed the effective dissemination of products. 

The QUBES platform currently hosts over 80 partner projects. These projects share a set of structural and technical needs that can burden the project team and undermine the time available to pursue the innovative project agendas. Hosting projects on the QUBES platform helps to avoid the inefficient and unsustainable practices of hosting work on a separate server. We have developed a project support system that makes it easy to establish communication, collaboration, and dissemination mechanisms both within the project team and the broader STEM education reform community. We provide a turnkey group space that can be customized to address specific project needs. Our support services are designed to address three common challenges groups face when hosting their project on QUBES: understanding the operation of the platform; maintaining a sense of ownership and branding; and adopting new collaboration and communication strategies to pursue their project. We also introduce planning resources that help the project leadership coordinate phases of their projects with the functionality they will require in their QUBES group space. These resources are available within a partner support group on QUBES for onboarding of new projects, which includes demonstrations of the effective use of the platform to support the creation, maintenance, and sustainability of project workspaces.

The CCBioInsites project is an example of an NSF funded project that hosted their activities on QUBES. The grant focused on helping community college faculty participate in discipline based education research and improving their teaching practices. The project recruited a distributed cohort of faculty and used their QUBES site to coordinate their activities \cite{Musgrove}.

Similarly the QB@CC project focused on teachers at 2-year schools who have limited access to professional learning opportunities. They have brought together both mathematics and biology faculty to collaboratively adapt existing teaching modules so that they reflect effective mathematics, biology content, and pedagogy. These products will persist and invite further customization in the QUBES OER Library disseminating and sustaining their efforts well beyond their active grant funding.
\vspace*{-5pt}

\subsection{Professional Learning: Faculty Mentoring Networks}
Professional learning refers to an intentionally designed collaborative environment where teachers work with one another to learn, develop and practice new methods for educating students. In contrast to some professional development models, professional learning promotes learning through engagement in reform practices, deemphasizing the role of telling faculty what to do. Providing effective and efficient professional learning experiences is essential to faculty participation in education reform. Equitable access to scholarly learning opportunities requires that those opportunities are not exclusively tied to events like conferences which can limit participation to those with available money and time. In STEM specifically, both the disciplinary knowledge base and the scientific tools are evolving rapidly, necessitating ongoing professional investment in teaching and learning. Access to professional learning is a major contributor to the implementation of evidence-based teaching practices discovered through discipline-based education research.

Our primary professional learning model is called a Faculty Mentoring Network (FMN). These involve geographically distributed groups of 10-15 faculty working together with a facilitator over the course of a semester. They use a QUBES platform group space to share resources, communicate asynchronously, and publish their products. The schedule and activities are structured to lead them through multiple stages: (1) a process of learning about a new teaching resource; (2) customizing that resource for use in their teaching setting and with their student audience; (3) implementing the module in their classroom; then (4) refining and publishing their adaptation to the OER Library. We have templated the processes necessary to run an FMN and provide training for FMN facilitators. Additionally, we have developed modules addressing common faculty interests such as inclusive teaching strategies, universal design for learning, and overcoming math anxiety which can be integrated across diverse FMNs.

The basic FMN model has been implemented in over 90 professional learning opportunities addressing diverse agendas. Groups have adapted our FMN model to address different outcomes including learning how to use new computational tools, designing new curricular activities, and conducting collaborative research. 

One example of how an FMN has been used to engage faculty in a scholarly approach to teaching and learning involves the extension and adaptation of a valuable teaching module. ``Investigating the footprint of climate change on phenology and ecological interactions in north-central North America'' was originally published as part of a NSF funded project coordinated by the Ecological Society of America \cite{Calinger}. An FMN was hosted on QUBES where faculty were mentored through customizing the original activity and teaching this data-intensive lesson as part of their professional learning experience. This FMN led to 15 published adaptations of the original activity where faculty produced customizations to incorporate different regional flora, use different analysis tools, and fit within different course contexts (\url{https://qubeshub.org/publications/267/forks/1}).
\vspace*{-5pt}

\subsection{Customizable Workspaces}
There are a wide range of professional communities (both formal and informal) that can play an important role in STEM education reform. In addition to the grant funded projects described above, communities involving professional societies, education nonprofits, research collaborations, and special interest groups can help engage faculty with professional opportunities. Programs such as conferences, webinars, and workshops can also help expose faculty to new ideas and potential collaborators. In order to be impactful these communities need to establish and sustain faculty engagement. Everything from cross-institutional course-based undergraduate research experiences (CUREs), to citizen science projects, to journal clubs and special interest groups can play a role in helping faculty pursue scholarship around teaching and learning. 

The QUBES platform currently hosts over 450 online group workspaces containing 1,200 project areas. We have a set of strategies for establishing, managing, and disseminating resources from distributed communities of faculty as they pursue new ideas for teaching and learning STEM. These involve integrating synchronous and asynchronous interactions, and using active facilitation techniques to work toward a shared community goal. Through the creation of customizable workspaces, the QUBES platform can be used to extend engagement with traditional face-to-face conferences as well as support hybrid and on-line only professional meetings. While participants can meet synchronously during meetings and workshops, the infrastructure also supports asynchronous engagement with material and other participants before and well after the synchronous portion of the meeting has ended.

As an example, BioQUEST runs an annual online meeting called the BIOME Institute which offers a unique opportunity to engage with a community of peers to address an educational challenge with the ultimate goal of improving student outcomes. A dedicated website and community space is created prior to the BIOME Institute. Participants are invited to this space ahead of the meeting to introduce themselves, bring attention to any educational reform projects they are involved in, and to engage with introductory prompts addressing the meeting focus and objective. During the meeting, talks and poster presentations are given throughout, working groups are brainstormed and formed, with all the material published as OER on the platform. Plans are made near the end of the synchronous portion of the meeting to create online sub-communities for working groups so that participants can continue their work asynchronously over the Fall semester. Products from these asynchronous working groups are then published as OER on the platform. Some of these working groups develop into grant collaborations. You can see a list of the recent working groups spun off from the BIOME meeting here (\url{https://qubeshub.org/community/groups/summer2021}).
\vspace*{-5pt}

\section{ONGOING AND FUTURE WORK}
In addition to the four platform services introduced above we continue to integrate new functionality and use scenarios to support STEM education reform. Since the inception of the QUBES platform in 2014, we have emphasized working closely with the user community to develop a shared vision for the ways that a gateway can support faculty scholarship around teaching and learning reflected in this quote.

\begin{quote}
If you want to build a ship, don't drum up people to collect wood and don't assign them tasks and work, but rather teach them to long for the endless immensity of the sea. -- Antoine de Saint Exupéry
\end{quote}

\begin{figure}
\centerline{\includegraphics[width=18.5pc]{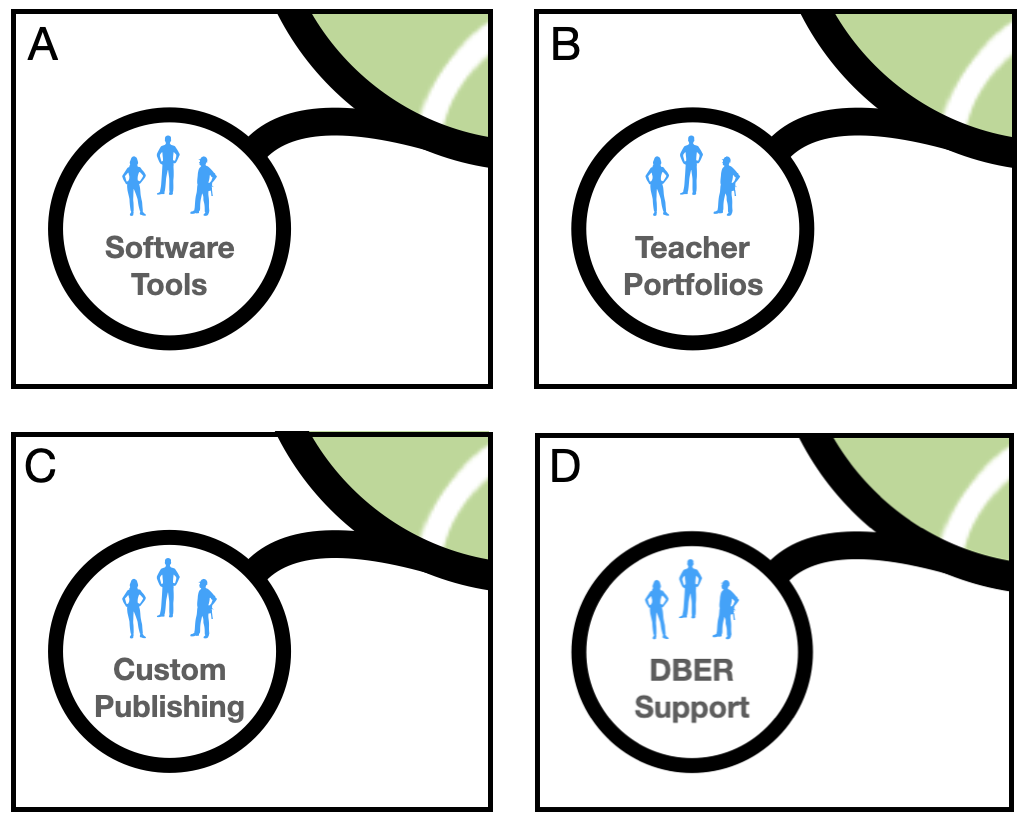}}
\caption{{\bf Example ongoing and future QUBES platform development projects.} Each of the four panels represent ongoing or future development to extend the functionality of the QUBES platform to further support faculty engagement in scholarship around teaching and learning. (A) Software tools involve implementing scalable and robust hosting of cloud based modeling and analysis tools in a way that is tightly integrated with OER published in the library. (B) Teacher portfolios refer to the design of an integrated system for collecting, sharing, and contextualizing faculty teaching scholarship as a means to document activities for consideration in hiring, promotion, and tenure. (C) Custom publishing involves streamlining the deployment of manuscript management workflows, and specialized publication collections. (D)
Discipline based education research support describes a suite of tools that would facilitate collaboration around classroom research and studies of reform practices.}
\label{fig:future}
\vspace*{-5pt}
\end{figure}

Our ongoing and future development plans for the platform stem directly from needs identified by hosted projects and users. Here we briefly introduce four high priority areas including: scalable and robust hosting of software tools, including tight integration of these tools with OER; design of teacher portfolios, akin to a LinkedIn or ResearchGate for educators; a custom publishing platform that supports peer-reviewed education journal tools, beyond the self-publishing of OER already supported on the platform; and support for discipline-based education research (DBER), which is constantly expanding our knowledge on evidence-based teaching strategies (see Figure \ref{fig:future}).

\subsection{Software Tools}
One of the key elements of a scientific gateway is simplified access to computational tools for research and teaching. The QUBES platform utilizes the open-source content management system HubZero in large part because of its built-in support for hosting computational tools, in addition to its collaboration and communication functionality. Addressing the scalability of access to these tools for use in the classroom has been challenging. Education user communities are much larger than research communities and require different types of server support. This issue was exemplified most recently by the COVID pandemic. During lockdown and social distancing, many biologists moved their research and teaching out of the in-vitro/in-vivo and into the in-silica, leading to an increased demand for simulations and software.

Data science has seen an explosion of demand and utility in the biological sciences, with a specific increase in students and faculty learning and using R and Python in their courses and research. There are many organizations working to train faculty in modern data science skills and techniques, such as The Carpentries, as well as offering professional development opportunities on how to incorporate these skills into the classroom. There are many challenges, however, in bringing these tools into the classroom. Posit Cloud, for example, has free access for students to RStudio without the need to install R and RStudio on their computers, but students can quickly run out of compute time, especially as they are learning how to code. CyVerse offers many services around accessing bioinformatics and computational tools in the cloud, but they mainly service the research community. We are actively exploring ways to increase accessibility to computational tools in classrooms, and fully integrate cloud based computing within the context of OER published in the library.

\subsection{Teaching Portfolios}
As faculty participate in reform projects, publish OER materials, engage with professional learning, and access materials from the OER library, they are building a record of their activity on the QUBES platform. Teaching scholarship is often difficult to document because there are few publishing outlets for teaching resources, with those that do get published tending to be cited less frequently as their use happens in the informal context of the classroom. Our OER publishing platform already provides DOIs, full citations, and some data about uptake by the community. Furthermore, each version and adaptation of a publication is tracked, showing iterative refinement, adoption, and use, with attribution automatically assigned on adaptation. Our goal is to supplement these impact metrics with additional ways for faculty to document their teaching scholarship. We plan to reconfigure the existing dashboards available to QUBES users to more effectively capture and share information about OER that has been published, groups joined, and FMNs completed. We will also extend the user profile module to capture more information related to faculty activities and backgrounds. 
 
We typically acknowledge successful completion of professional learning opportunities through FMNs and workshop/meeting experiences with certificates and letters of participation. In the future we will establish a badge system to document a wide range of activities that faculty complete. These badges can be linked not just to learning opportunities completed, but also to products produced via those opportunities. Essentially, our vision for teaching portfolios involves providing more flexible opportunities to organize, present, and annotate one’s contributions. We are already seeing many faculty, particularly those in non-R1 settings, using information from their profiles and dashboards to capture their scholarship in annual activity reports.

\subsection{Custom Publishing}
The vast majority of QUBES OER are submitted as a ``QUBES Resource.'' This resource type contains a broad set of metadata for describing the resources, their intended audience, and other features like their use of inclusive learning practices, universal design principles followed, and racial equity strategies employed. QUBES currently provides custom publishing types for a small group of projects including Math Modeling Hub (\url{https://mmhub.qubeshub.org}), NIBLSE (\url{https://niblse.qubeshub.org}), CourseSource (\url{https://coursesource.qubeshub.org}), and SIMIODE (\url{https://simiode.qubeshub.org}). These custom resource types have their own metadata schema and can be set to employ a review process before public release. For example, the NIBLSE project developed and published a set of core competencies related to describing beneficial outcomes of integrating bioinformatics techniques throughout the biology curriculum. These competencies are represented in their metadata schema, thereby raising awareness and improving findability across their collection of publications. CourceSource, a peer reviewed journal of biology and physics teaching resources, transferred their OER library into QUBES, where we developed a custom metadata schema including alignment with learning outcomes and teaching frameworks. In collaboration with CourseSource, we are currently building a custom submission and editorial management pipeline that will allow authors, editors, and reviewers to utilize QUBES throughout the submission, review, and publication process. 

The goal is to make the OER manuscript management process customizable for easier and cost effective utilization by other communities. We see implications for ingesting existing education libraries looking for a sustainable home, and support for distributed communities, such as CUREs and citizen science projects. Custom designed metadata schemas and curation pipelines will support individualized, autonomous search-and-browse portals within their communities, with their resources also available across the entire QUBES OER Library for broad dissemination and discoverability.

\subsection{Discipline Based Education Research Support}
There is growing awareness of and focus on discipline based education research as an important component of faculty scholarship. We imagine QUBES as a platform where curriculum specialists, education researchers, and teaching faculty could collaborate to scale up data collection and explore the impacts of interventions in diverse teaching contexts and across student audiences. This would likely involve some tools to facilitate data collection and management (e.g., surveys and other online forms, activity tracking, and use logging). We would also need to institute student account types that would appropriately handle data de-identification to mitigate any personal risks. QUBES has already been used to coordinate distributed research projects using faculty mentoring networks to connect researchers with motivated teachers. We believe that the gateway environment can play an important role facilitating research on faculty change, professional learning, project management, as well as documenting emerging practices as communities adopt gateway infrastructures to evolve their professional practices.

\section{MOVING FORWARD - A CALL TO ACTION}
The QUBES platform has been designed and implemented to help faculty pursue scholarship around teaching and learning. Both the technical infrastructure (tools) and social infrastructure (practices) were purpose built to advance innovation in STEM education by supporting communities of practice, foregrounding equity, diversity and inclusion, and engaging faculty in the culture of open education. Our four platform services (OER Library Access; Professional Learning; Partner Support; and Customizable Workspaces) provide multiple points for faculty engagement and address key aspects of accelerating reform practices. Hosting diverse activities on a single platform creates a synergistic effect and has proven to be important to our success, allowing us to capture faculty work and make it accessible to the broader community. We currently host a community of over 20,000 registered users and we are beginning to see the impact of both  leveraging the platform to get work done and the ways that feeds forward into the collection, curation, and documentation of those professional activities. The integration and interoperability of the various gateway tools supports and contextualizes a broad network of projects, organizations, and faculty. 

We believe that the broad use of the QUBES platform by multiple stakeholders demonstrates a significant need for a centralized infrastructure to support reform practices. The National Science Foundation has identified QUBES, along with other emerging science education gateways, as important resources for increasing sustainability and broadening the impact of funded education projects \cite{NSFDCL}. We encourage further exploration of the ways that science education gateways might prove to be fruitful across STEM education, helping to develop other purpose-built communities and platforms to accelerate innovation in teaching and learning science.
\vspace*{-8pt}

\section{ACKNOWLEDGMENTS}
This material is based upon work supported by the National Science Foundation under Grants  1346584, 1446258, 1446269, 1602989, and 1446284. The authors would like to thank HubZero and the Science Gateways Community Institute for their support.

\def\refname{REFERENCES}

\vspace*{-8pt}

\end{document}